\documentclass{aa}  

\usepackage{graphicx}
\usepackage{txfonts}
\usepackage{amssymb}
\usepackage{amsmath}
\usepackage{graphicx}
\usepackage[colorlinks=true, allcolors=blue]{hyperref}
\usepackage{booktabs}
\usepackage{amsfonts}     
\usepackage[T1]{fontenc} 
\usepackage[utf8]{inputenc} 
\usepackage{multicol}
\usepackage{xcolor}
\usepackage{lineno}
\usepackage{gensymb}
\usepackage{soul}
\usepackage{rotating}
\usepackage{comment}

\begin{document}

   \title{The polarisation behaviour of OJ 287 viewed through radio, millimetre and optical observations between 2015 and 2017}

   \subtitle{}

   \author{J. Jormanainen
          \inst{1}\fnmsep\inst{2}\fnmsep\inst{3}\fnmsep\thanks{jesojo@utu.fi}\and
          T. Hovatta\inst{1}\fnmsep\inst{3}\and
          E. Lindfors\inst{2}\and
          A. Berdyugin\inst{2}\and
          W. Chamani\inst{4}\and
          V. Fallah Ramazani\inst{1}\fnmsep\inst{3}\and
          H. Jermak\inst{5}\and
          S. G. Jorstad\inst{6}\fnmsep\inst{7}\and
          A. Lähteenmäki\inst{3}\fnmsep\inst{8}\and
          C. McCall\inst{5}\and
          K. Nilsson\inst{1}\and
          P. Smith\inst{9}\and
          I. A. Steele\inst{5}\and
          J. Tammi\inst{3}\and
          M. Tornikoski\inst{3}\and
          F. Wierda\inst{2}
          }

   \institute{Finnish Centre for Astronomy with ESO, University of Turku, Finland
         \and
             Department of Physics and Astronomy, University of Turku, FI-20014, Finland
         \and
            Aalto University Metsähovi Radio Observatory, Metsähovintie 114, FI-02540 Kylmälä, Finland
        \and
            Helmholtz centre Potsdam, GFZ German Research Centre for Geosciences, Germany
        \and
            Astrophysics Research Institute, Liverpool John Moores University, Liverpool Science Park IC2, 146 Brownlow Hill, Liverpool L3 5RF, UK
        \and
            Institute for Astrophysical Research, Boston University, 725 Commonwealth Avenue, Boston, MA 02215, USA
        \and
            Saint Petersburg State University, 7/9 Universitetskaya nab., 199034 St. Petersburg, Russia
        \and
            Aalto University Department of Electronics and Nanoengineering, P.O. BOX 15500, FI-00076 AALTO, Finland.
        \and
            Steward Observatory, University of Arizona, 933 N. Cherry Ave., Tucson, AZ 85721 USA
             }

   \date{Received xxx; accepted yyy}

 
  \abstract{OJ 287 is a bright blazar with century-long observations, and one of the strongest candidates to host a supermassive black hole binary. Its polarisation behaviour between 2015 and 2017 (MJD 57300-58000) contains several interesting events that we re-contextualise in this study. We collected optical photometric and polarimetric data from several telescopes and obtained high-cadence light curves from this period. In the radio band, we collected mm-wavelength polarisation data from the AMAPOLA program. We combined these with existing multifrequency polarimetric radio results and the results of very-long-baseline-interferometry imaging with the Global mm-VLBI Array at 86 GHz. In December 2015, an optical flare was seen according to the general relativistic binary black hole model. We suggest that the overall activity near the accretion disk and the jet base during this time may be connected to the onset of a new moving component K seen in the jet in March 2017. With the additional optical data, we find a fast polarisation angle rotation of $\sim 210\degree$ coinciding with the December 2015 flare, hinting at a possible link between these events. Based on the 86-GHz images, we calculated a new speed of 0.12 mas/yr for K, which places it inside the core at the time of the 2015 flare. This speed also supports the scenario where the passage of K through the quasi-stationary feature S1 could have been the trigger for the very-high-energy gamma-ray flare of OJ 287 seen in February 2017. With the mm-polarisation data, we established that these bands follow the cm-band data but show a difference during the time of K passing through S1. This indicates that the mm-bands trace the substructures of the jet still unresolved in the cm-bands.
  }


   \titlerunning{Multiwavelength view of the long 2016 EVPA rotation of OJ 287}
   \authorrunning{Jormanainen et al.}

   \maketitle
%

\section{Introduction}
\label{intro}

Blazars are active galactic nuclei (AGN) where we see relativistic jets shooting out from the poles of their central supermassive black hole (SMBH), pointing at a very small inclination with respect to our line of sight \citep{Blandford_Rees1978,Urry_Padovani1995}. This geometry and the relativistic jet speeds greatly boost their perceived luminosity and variability time scales. This is also why most of the observed emission that we see extending across the entire electromagnetic spectrum originates in the jet \citep{Scarpa1997,Blandford2019,Hovatta2019}. The spectral energy distribution (SED) of blazars possesses a typical two-bump structure, where the first, lower energy bump is known to originate from the synchrotron emission. The exact mechanism of the higher energy bump is not known but it is thought to be dominated by the inverse Compton emission \citep{Rees_1967MNRAS.135..345R}. The synchrotron emission typically extends from the radio bands sometimes all the way up to X-rays, and is highly variable in flux but also in polarisation. Studying the polarisation signatures in particular allows the study of the source magnetic field, the jet geometry, and acceleration mechanisms.

OJ 287 is a bright and nearby blazar that is also one of the most promising candidates to host a binary SMBH \citep{Valtonen_2008Natur.452..851V,Valtonen2016ApJ...819L..37V}. Its optical variability shows quasi-periodic trends of 12 and 60 years \citep{Sillanpaa_1996A&A...305L..17S, Valtonen_2006ApJ...646...36V}. Several models have been proposed to explain its variability behaviour in the optical band, typically invoking the secondary black hole impacting the accretion disk of the primary black hole \citep{Valtonen_2008Natur.452..851V}, jet precession caused by the secondary \citep{Dey2021MNRAS.503.4400D}, wobbling jet \citep{Agudo_2012ApJ...747...63A}, but also precession \citep{Liska_2018MNRAS.474L..81L,Britzen_2018MNRAS.478.3199B,Britzen2023ApJ...951..106B} and shock-in-jet \citep{Marscher2008Natur.452..966M} models that require no binary system. The optical polarisation properties of OJ 287 have also been studied in detail \citep[e.g.][]{Holmes_1984MNRAS.211..497H,Sillanpaa_1991AJ....101.2017S,Villforth2010MNRAS.402.2087V,Blinov_2011ARep...55.1078B,Gupta2019AJ....157...95G,Gupta2023ApJ...957L..11G} and polarisation studies could help further confirm this binary nature. The quasi-periodic flares caused by an impact of a secondary BH on the accretion disk of the primary BH should have a thermal origin \citep{Lehto_Valtonen_1996ApJ...460..207L} and thus a low polarisation degree during the flare would act as a tell-tale signature for this \citep[e.g.][]{Smith_1985AJ.....90.1184S,Valtonen_2008Natur.452..851V,Valtonen2016ApJ...819L..37V,Villforth2010MNRAS.402.2087V,Valtonen_Sillanpaa_201120052010multiwavelength}. Furthermore, observations of total flux and polarisation in the radio band band can help distinguish between the different models.  

Between December 2015 and January 2017 (MJD 57370–57785) OJ 287 was seen undergoing a long, persistent electric vector polarisation angle (EVPA) rotation for a total of $\sim340\degree$ ('2016 event' from here on) in several radio bands \citep{Myserlis2018A&A...619A..88M}. To explain this, the authors considered two different model types, namely multicomponent models \citep[e.g.][]{Bjornsson_1982ApJ...260..855B,Sillanpaa_1992A&A...254L..33S,Villforth2010MNRAS.402.2087V,Cohen2018ApJ...862....1C} and models with a curved trajectory. In the latter type of models, they also made the division of the curved trajectory into two different kinds of motion: motion of an emission component down a helical magnetic field \citep{Marscher2008Natur.452..966M} and motion down a bent jet \citep[e.g.][]{Gomez_1994A&A...284...51G,Abdo2010Natur.463..919A}. The slow, gradual rotation that was seen in the radio bands, starting at different times for the 10.45 and 8.35 GHz bands and much later for the 4.85 GHz band, was attributed to the motion of the emission component down a bent jet, the differences in the starting time resulting from the effect of optical depth. The jet of OJ 287 has been imaged with the space-based very-long-baseline-interferometer (VLBI) telescope RadioAstron at 1.7 GHz \citep{Cho2024A&A...683A.248C}, with RadioAstron at 22 GHz and with the Global mm-VLBI Array (GMVA) at 86 GHz \citep{Gomez2022ApJ...924..122G}, as well as with GMVA + phased ALMA at 86 GHz \citep{Zhao2022ApJ...932...72Z}, and the bend is clearly resolved in these images. In \cite{Myserlis2018A&A...619A..88M}, the more erratic optical EVPA showed faster subrotations on top of a similar declining trend as the radio data, and this was thought to be related to the helical motion of the emission component down the source magnetic field, although the optical V-band data had long gaps. Synchronous variability of the optical and radio polarisation has been observed in the past from OJ 287 by \citep{Kikuchi_1988A&A...190L...8K} where they suggest a connection between the optical and radio emitting regions and a common underlying mechanism.

On the other hand, \cite{Cohen_Savolainen2020A&A...636A..79C}, demonstrated that with a simple two-component model \citep{Cohen2018ApJ...862....1C} they could explain several types of rotations in blazars, including the 2016 event seen in OJ 287. In this work, they demonstrated how different types of rotations within the same source can results from the relative flux and EVPA ratios of a stable jet component and an outburst component, the presence or absence of a rotation thus being a phase effect. In the case of the 2016 event rotation, they showed that only a 10$\degree$ difference is needed between the jet and the burst EVPA to cause the long rotation and that a slightly different burst flux between the 10.45 and 8.35 GHz data also explains the different behaviour, whether the rotation is seen clearly or not. Physical counterparts for multicomponent models have also been suggested via simulations such as in \cite{Nakamura_2010ApJ...721.1783N,Nakamura_Meier2014ApJ...785..152N,Zhang_2014ApJ...789...66Z,Zhang_2015ApJ...804...58Z}.

\cite{Lico2022A&A...658L..10L} reported of their observations of OJ 287 with the GMVA and phased ALMA at 86 GHz between May 2015 and September 2017, coinciding with the time of the 2016 event. In the VLBI images that they modelled, they distinguished a moving component (knot) K between two quasi-stationary features S1 and S2 \citep[S1 and S2 also identified in][]{Zhao2022ApJ...932...72Z} in March 2017. In their study, they presented a compelling possible link between the passage of K through the quasi-stationary feature S1 and the first very-high-energy (VHE, E > 100 GeV) gamma-ray detection of OJ 287 \citep{Mukherjee2017ATel10051....1M}.

In this paper, we reinterpret the 2016 event and consider the propagation of the moving component identified in \cite{Lico2022A&A...658L..10L} in the context of the data used in \cite{Myserlis2018A&A...619A..88M} together with additional optical polarisation data and mm-range data from the ALMA polarisation monitoring program AMAPOLA \citep{Kameno_2023atyp.confE..38K}. 
This paper is structured as follows. In Sect. \ref{data}, we describe the data included in this study. In Sect. \ref{analysis} we describe our analysis process and the results obtained. Finally, in Sect. \ref{discussion}, we discuss our results in the context of previous studies, how our data can be used to refine these, and give our conclusions.

\section{Data}
\label{data}

For all bands, we resolved the $180\degree$ ambiguity of the EVPA by following \cite{Kiehlmann_2016A&A...590A..10K} where each subsequent data point is adjusted according to

\begin{align}
    \chi _{adj,i} = \chi _{i}-n\pi, \ \mathrm{where} \ n = \mathrm{int}\Bigg(\dfrac{\chi _{i}-\chi _{i-1}}{\pi}\Bigg).
\end{align}

\noindent Here $\chi _{i}$ is the observed EVPA value and $\chi _{adj,i}$ the adjusted EVPA value. The notation int($\cdot$) means rounding to the nearest integer. We corrected both the optical and the radio EVPA in this manner so that the EVPA is seen as a continuous curve (see Figs. \ref{fig:rad_timings} and \ref{fig:opt_timings}). We treated each radio band separately but the optical V-, R-, and g*-bands together as a single data set since the EVPA in these bands display negligible differences (see Sect. \ref{data:optical}).

\subsection{Radio}
\label{data:radio}

In this work, we use some of the multiwavelength (MWL) total flux and linear polarisation data from \cite{Myserlis2018A&A...619A..88M}. These include the data from 2.64, 4.85, 8.35, and 10.45 GHz bands taken with the Effelsberg 100-m telescope as part of a large multifrequency campaign \citep{Komossa_2015ATel.8411....1K} started in December 2015 (MJD 57370). The reduction process of the Effelsberg data are described in \cite{Myserlis2018A&A...619A..88M}.

To put the 2016 event described in \cite{Myserlis2018A&A...619A..88M} in context in a more complete MWL picture, we extended the period to span between October 5, 2015 and September 4, 2017 (MJD 57300-58000). We collected data from the well-sampled Metsähovi blazar monitoring program at 37 GHz from this entire period. The reduction process of the Metsähovi data is described in \cite{Terasranta_1998A&AS..132..305T}.

\begin{figure*}
\centering
\includegraphics[scale=0.5]{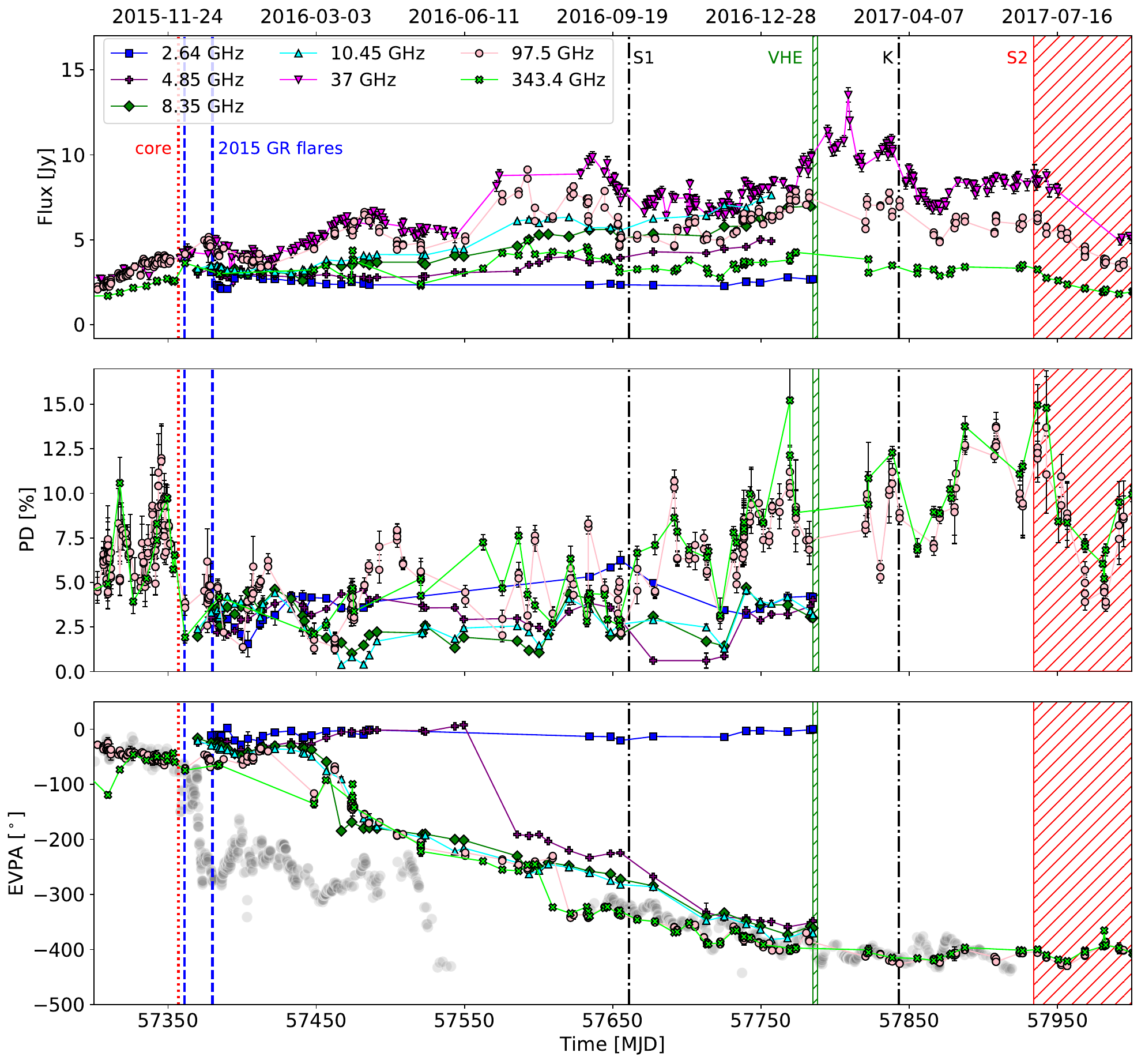}
  \caption{Light curves displaying the multifrequency radio flux, PD and EVPA evolution. Starting from MJD 57586, the EVPA data at 4.85 GHz has been shifted by $180\degree$ for better comparison with the higher frequencies. The grey data points in the bottom panel show the evolution of the optical data (shown in Fig. \ref{fig:opt_timings}) for contrast. The blue dashed vertical lines mark the times of the optical flares predicted by the GR-model. The black dot-dashed vertical lines mark the GMVA epochs when the quasi-stationary feature S1 was seen brightened and when the moving component K was detected. The green hatched area marks the period of VHE flaring. The red vertical dotted line marks the time when K should have been still residing within the mm-radio band core, and the red hatched area the estimated time when K has reached the quasi-stationary feature S2 (see details in Sects. \ref{analysis} and \ref{discussion}).}
  \label{fig:rad_timings}
\end{figure*}

From the same extended period, we collected data from the ALMA polarisation monitoring program AMAPOLA\footnote{\url{https://www.alma.cl/~skameno/AMAPOLA/}} to bridge the gap between the lower radio frequencies and the optical data. We added the data from Bands 3 (97.5 GHz) and 7 (343.4 GHz), which had adequate sampling. The AMAPOLA program is described in \cite{Kameno_2023atyp.confE..38K}. For the ALMA bands, we removed those polarisation degree (PD) and EVPA points whose corresponding significance of the polarised flux density(PF) was less than $3\sigma$. For Band 3 the removed data was $\sim 15\%$ of the data and for Band 7 $\sim 3\%$. We show the evolution of flux density, PD, and EVPA in Fig. \ref{fig:rad_timings}.

\subsection{Optical}
\label{data:optical}

In the optical band, \cite{Myserlis2018A&A...619A..88M} used V-band polarisation data of the Steward Observatory blazar monitoring program\footnote{\url{https://james.as.arizona.edu/~psmith/Fermi/}}. As described in Sect. \ref{data:radio}, we expanded the observation period between MJD 57300-58000 and included more data from the the Steward Observatory accordingly. We collected optical data from other instruments to increase the sampling and especially to trace the EVPA behaviour in detail. These include R-band photometry from the Steward Observatory, R-band polarimetry and photometry data from the Boston University Blazar monitoring\footnote{\url{https://www.bu.edu/blazars/BEAM-ME.html}}, g*-band polarimetry data from the RINGO3 \citep{Arnold2012SPIE.8446E..2JA} monitoring program, and V- and R-band polarimetry data from DIPOL-2 polarimeter \citep{Piirola_2014SPIE.9147E..8IP}. The Steward Observatory reduction process is described in \citet{Smith2009arXiv0912.3621S}, the Boston University program in \cite{Jorstad_Marscher_galaxies4040047}, and the RINGO3 program in \cite{MCCall_2024MNRAS.532.2788M}. We included R-band photometry and polarimetry data from the St. Petersburg observatory in Russia, Crimean observatory, as well as from the Kanata telescope in Hiroshima, Japan published in \cite{Gupta2023ApJ...957L..11G}. Additionally, we included optical R-band photometry data from the Tuorla blazar monitoring program\footnote{\url{https://users.utu.fi/kani/1m/}} \citep{Takalo_2008AIPC.1085..705T}, the analysis of which is described in \cite{Nilsson_2018A&A...620A.185N}.

All PD data were treated for the Ricean bias using $p_{debiased} \approx \sqrt{p^{2}-\sigma _{p}^{2}}$, where $p$ is the measured polarisation fraction and $\sigma _{p}$ its error, for values with $p/\sigma _{p} < 3$ \citep{Vaillancourt2006PASP..118.1340V,Pavlidou_2014MNRAS.442.1693P}. For all optical data, we removed those PD and corresponding EVPA data points for which the significance of the PD was less than $3\sigma$. The removed data points were $\sim 2\%$ of all data available during this period.

\begin{figure*}
\centering
\includegraphics[scale=0.5]{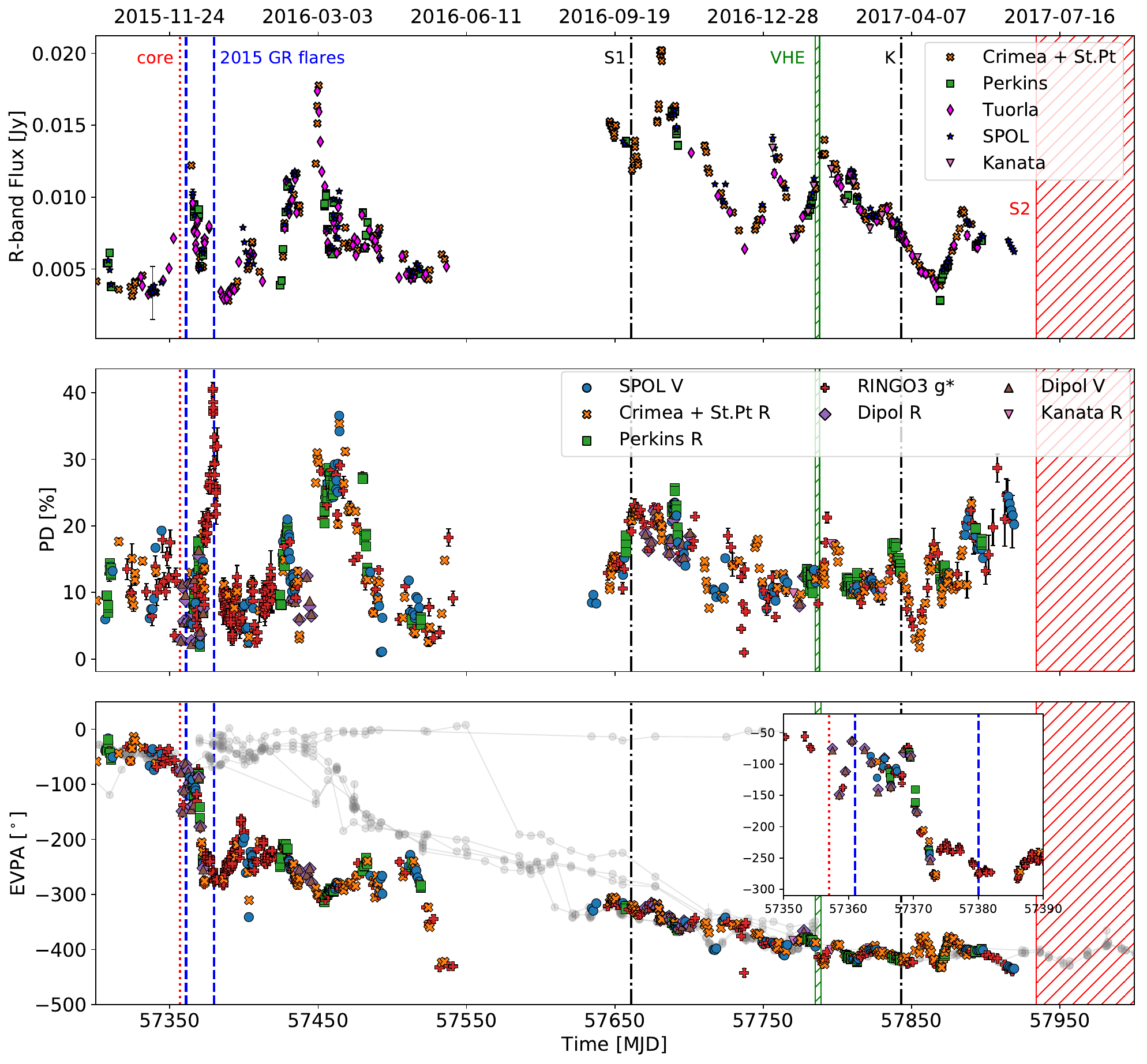}
  \caption{Light curves displaying the R-band flux, and combined R-, V- and g*-band PD and EVPA evolution. The inset shows a zoom-in of the $\sim 210\degree$ rotation seen in the optical data. The grey data points in the bottom panel show the evolution of the radio band (as shown in Fig. \ref{fig:rad_timings}) for contrast. The vertical lines and hatched areas are as explained in Fig. \ref{fig:rad_timings} (see details in  Sects. \ref{analysis} and \ref{discussion}).}
  \label{fig:opt_timings}
\end{figure*}

Because we combined the data from R- and V-bands as well as from the g*-band \citep[that is the closest equivalent to R-band from the three RINGO3 bands, see][]{MCCall_2024MNRAS.532.2788M}, we cross-checked the simultaneous values from these bands. The differences in both PD and EVPA were small enough (EVPAs within 1-2 degrees) not to affect any of our conclusions. The optical fluxes, PD and EVPA evolution are shown in Fig. \ref{fig:opt_timings}.

\section{Analysis and results}
\label{analysis}


In the radio regime, we see flares with increasing amplitudes most prominently described in the 97.5 GHz band and especially in the 37 GHz intermediate band. The highest peak of flux in the 37 GHz band is around MJD 57800 with 15.5 Jy. At the time of this flare, the 97.5 GHz band showed elevated activity. The second highest peak is at around MJD 57640 at 37 GHz with 10.0 Jy. The flux varies relatively slowly, showing about five roughly defined peaks during the two-year period. During this time, the PD at 97.5 and 343.4 GHz changes erratically, the highest peaks seen in the beginning of the period, around MJD 57350 with 12$\%$ at 97.5 GHz and towards the end of the period, around MJD 57770 with 15.0$\%$ at 343.4 GHz. After this the PD frequently reaches above 10.0$\%$ in both bands, once more reaching  15.0$\%$ around MJD 57940. The evolution of the EVPA at 97.5 and 343.4 GHz appears to follow a similar slow rotational trend revealed in \cite{Myserlis2018A&A...619A..88M} in the lower frequencies. In the bottom panel of Fig. \ref{fig:rad_timings}, starting from MJD 57586, we have shifted the 4.85 GHz band by $-180\degree$ for an easier comparison of the rotation behaviour in the higher bands. As we can see, the EVPA evolutions between the cm- and mm-bands differ most clearly before and during the faster part of the rotation at 10.45 GHz around MJD 57450 and later between MJD 57570 and 57730.

In the optical band, the important note to make is the occurrence of a double-peaked flare, the first one peaking on December 5, 2015 (MJD 57361) and the second one seen on December 24, 2015 (MJD 57380). This flare was predicted and observed according to the general relativistic (GR) binary black hole model \citep{Valtonen_2011ApJ...742...22V, Valtonen2016ApJ...819L..37V}, and we mark the times of both peaks of the outburst in blue vertical dashed lines in Figs. \ref{fig:rad_timings} and \ref{fig:opt_timings} based on the timing of the flares in that work. The optical R-band photometric data available here (see the top panel of Fig. \ref{fig:opt_timings}) do not fully sample the flare that was seen reaching 21 mJy (corresponding to 12.9 magnitudes) in the better sampled data set shown in \citep{Valtonen2016ApJ...819L..37V}. In addition to the double flares, the flux evolution shows other nearly as high peaks on MJD 57450 with 18 mJy, on MJD 57680 with 20 mJy. The optical PD shows fast and erratic variability, reaching its highest peak of 40$\%$ around MJD 57380, the second highest peak of 36.5$\%$ around MJD 57460 and often reaching 20$\%$ and above after these. According to the GR model, the first flare of the December 2015 double flare should result from the secondary BH impacting the accretion disk of the primary BH. Thus, the origin of the first flare should be thermal, resulting in unpolarised emission. 
At the time of these flares, the PD is first at a minimum of 5\% during the first peak. In contrast, the second peak has a high of 40\% \citep[as reported in][]{Valtonen2016ApJ...819L..37V}, suggesting a synchrotron origin instead. With the improved cadence of the optical polarimetry data, the EVPA curve follows the same pattern already visible in \cite{Myserlis2018A&A...619A..88M}, with a first optical rotation seen starting around MJD 57370 for a total of $\sim 210\degree$, and several smaller rotations following this event. The coincidence of the fast EVPA rotation with the GR flare could suggest that the disk impact causes some changes in overall activity near the accretion disk and the jet base such that new material is launched down the jet, which we see as the rotation. Such a fast rotation is not traced in any of the radio bands where only a long, moderate rotation trend is observed. This moderate rotation is also visible in the optical band after the seasonal gap, MJD 57630 onward.

We detail our results in the following subsections. First, we describe the analysis of the EVPA data both in the radio and in the optical bands, highlighting the importance of the observing cadence with regards to the EVPA rotations. We also employ the the imaging results of \cite{Lico2022A&A...658L..10L} and use their results in a more detailed analysis of the moving component speed.

\subsection{Radio and optical EVPA}
\label{analysis:evpa}

The optical V-band data shown in \cite{Myserlis2018A&A...619A..88M} contain very long gaps, and, for blazars, it has been demonstrated \citep{Kiehlmann2021MNRAS.507..225K} that typically at least a 1-day cadence is necessary to identify rotational events when the rotation rate is fast. Therefore, the deductions based on the optical data leave room for interpretation, although there have been several studies showing evidence for the existence of a toroidal/helical magnetic field in the jet of OJ 287 \citep{Gomez2022ApJ...924..122G, Cho2024A&A...683A.248C, Lico2022A&A...658L..10L}, which \cite{Myserlis2018A&A...619A..88M} suggest as the cause of the subrotations seen in the optical EVPA. We collected much more data in the optical V-, R- and g*-bands to get a clearer picture, especially knowing that the variations in OJ 287 are often fast \citep[e.g.][]{Gupta2023ApJ...957L..11G}, and corrected these data as described above.

After correction, the mm-bands seem to follow the lower frequency radio bands rather well as seen in the bottom panel of Fig. \ref{fig:rad_timings}. Contrary to the data shown in \cite{Myserlis2018A&A...619A..88M}, we shifted the 4.85 GHz data by $-180\degree$ after MJD 57550 to confirm the similarity of the rotation trend between each band. With the additional optical data, we can see that while the same general structure seen already in \cite{Myserlis2018A&A...619A..88M} still exists we see much more substructure in the EVPA. The first fast rotation of $\sim 210\degree$, starting around MJD 57370, has a rotation rate of $-49\degree/\mathrm{d}$ (within 5 days). The fastest portion of the rotation has a sampling cadence of <1.0 days, confirming its accuracy. 
\cite{Myserlis2018A&A...619A..88M} found a rotation rate of $-0.7\degree/\mathrm{d}$ for the 8.35 and 10.45 GHz bands between MJD 57500 and 57780. After this the EVPA became constant with $\sim 0\degree/\mathrm{d}$. We confirmed similar rates with both 97.5 and 343.4 GHz with $-0.78\degree/\mathrm{d}$ and $-0.73\degree/\mathrm{d}$, respectively. After the seasonal gap, around MJD 57630-57780, the rotation rate was similar also in the optical band $-0.60\degree/\mathrm{d}$. All bands seem to stabilise after MJD 57780. The stable EVPA is also seen in all bands prior to the start of the rotation events, in radio bands before MJD 57430 and in optical before MJD 57350, the optical EVPA behaviour being rather erratic still. Notably, all the radio bands and the optical band rotate approximately one full rotation starting from $\sim-50\degree$ and stabilising at $\sim-400\degree$. 

As we discussed above, a good observing cadence is crucial for correctly identifying EVPA rotational events. We know that in the optical band the fastest changes in photometry and polarimetry can be seen on intra-day timescales \citep[seen in OJ 287, e.g. in][]{Sillanpaa_1991AJ....101.2017S,Gupta2017MNRAS.465.4423G,Gupta2019AJ....157...95G}, whereas in the radio band, the flux density often evolves more slowly, and we could expect the same behaviour to be true for its polarisation. Therefore, with the faster changes taking place in the optical band, the cadence plays a more important role. Because of the $180\degree$ ambiguity, changes larger than $90\degree$ cannot be traced between subsequent data points. This means that when the intrinsic EVPA shows changes faster than the observing cadence allows, the adjustment of data points to only show changes smaller than $90\degree$ misinterprets the intrinsic evolution \citep{Kiehlmann2021MNRAS.507..225K}. The EVPA curve in the mm-bands contains long gaps, and especially during the time or shortly after the fast optical rotation there are gaps as long as 31 days (MJD 57448-57417) in the 97.5 GHz and 72 days (MJD 57456-57384) in the 343.4 GHz. Therefore, with the relatively fast rotation rate of $-4\degree/\mathrm{d}$ in the 10.45 GHz band during the period MJD 57450-57500, a similar rotation in the mm-bands could have gone unnoticed. 

\subsection{86 GHz moving component speed}
\label{analysis:speed}

The analysis of the 86 GHz VLBI images in \cite{Lico2022A&A...658L..10L} revealed a detailed and an evolving jet structure of OJ 287. In their study, they modelled the source brightness distribution and show the total intensity images of five epochs between May 2015 and September 2017 as well as their respective flux densities \citep[see Figs. 1 and 2 in][]{Lico2022A&A...658L..10L}. Specifically, the three last epochs showed first the brightening of a quasi-stationary feature S1 over the entire image, a newly emerged moving component K between the S1 and another, further out quasi-stationary feature S2, and lastly the brightening of S2 feature and the K component having disappeared. The moving component K was detected in these images during MJD 57843 (March 2017). Quasi-stationary features S1 and S2 were seen dominating the total intensity at times MJD 57661 (September 2016) and  MJD 58025 (September 2017), and in \cite{Lico2022A&A...658L..10L} they interpret this as an indication of K passing through S1 and, subsequently S2, K thus being blended within these features. We mark two GMVA epochs, the S1 seen brightened and the detection of K, in Figs. \ref{fig:rad_timings} and \ref{fig:opt_timings} in black vertical dot-dashed lines.

An important aspect of their study was the first VHE gamma-ray detection of OJ 287 with the VERITAS telescopes very close to the detection of the moving component K. The observations that were initiated due to a high state in the soft X-rays (0.3-10 keV) seen with \textit{Swift}-XRT, showed a preliminary detection with a significance of 5.7$\sigma$ between February 1-4, 2017 (MJD 57785-57788) \citep{Mukherjee2017ATel10051....1M}. Later, further analysis of the VHE high state period between December 9, 2016 and March 31, 2017 (MJD 57731-5784) lead to an increased significance of $10\sigma$ of the detection \citep{Acharyya2024arXiv240711848A} and pinpointed the flaring epoch to February 1-5, 2017 (MJD 57785-57789, green hatched region in Figs. \ref{fig:rad_timings} and \ref{fig:opt_timings}). 


Based on the 86 GHz images and the variations in the component fluxes, \cite{Lico2022A&A...658L..10L} derived a speed for the moving component K. They assumed that K is still blended in with S1 on MJD 57785 at the time of the VHE high state, and used this date as the starting date for their speed estimation. Measured from the core, the average distance of S1, and thus also the distance of K at the time of the VHE flare, was $\sim 0.1$ mas. K was first detected between S1 and S2 on MJD 57843, at $\sim 0.16$ mas. Thus, using the VHE high state as their starting date and the detection as their ending date, they obtained the speed of 0.38 mas/yr. However, they also stated that K must have been within S1 already on MJD 57661 when the flux of S1 was dominating the total intensity. Placing K at the distance of $\sim 0.1$ mas already at this time and using the same ending time and distance as in \cite{Lico2022A&A...658L..10L}, we can obtain a new, slower speed of 0.12 mas/yr for K. With this speed, the time for K to reach S2 matches well with the epoch MJD 58025 when the flux of S2 is seen dominating. We mark the arrival epoch of K to S2 in red hatched area in Figs. \ref{fig:rad_timings} and \ref{fig:opt_timings} as there is some uncertainty in the position of S2 and thus the arrival time of K. Knowing that S1 is at ~0.1 mas, we also calculated back the time when K still must have been inside the 86 GHz core. Our estimated time is around MJD 57357 (red vertical dotted line in Figs. \ref{fig:rad_timings} and \ref{fig:opt_timings}), which coincides with the start of the GR flare, as we discuss in the next section.

Similar to what was done in \cite{MAGIC_2018A&A...619A..45M}, we calculated the time it takes for K to cross S1 by using the estimated component sizes determined by \cite{Lico2022A&A...658L..10L} (see their Table 1). The average size of S1 across all their measured epochs is $(0.037 \pm 0.021)$ mas. Thus, with the faster speed obtained by \cite{Lico2022A&A...658L..10L}, the passage would only take $(35 \pm 20)$ days. The time between the S1 brightening (MJD 57661) and the VHE detection (MJD 57785-57789) is 124 days and with this faster speed it seems unlikely that K could still have interacted with S1 at the time of the VHE flare. Using our slower speed, the time it takes for K to pass through S1 is $(112 \pm 64)$ days, making this scenario more likely.

Further evidence for the slower speed we calculated above is provided by the MOJAVE\footnote{\url{https://www.cv.nrao.edu/MOJAVE/}} data at 15 GHz \citep[][]{Lister2021ApJ...923...30L}. In 2017 and 2018, two components are resolved at 0.2 mas (ID 30) and 0.4 mas (ID 29) distance from the core with speeds 0.13 and 0.16 mas/yr, respectively. One of these could correspond to component K that becomes visible later at the lower frequency images with poorer angular resolution, matching the slower speed we determined for K.


\section{Discussion and conclusions}
\label{discussion}

The behaviour of OJ 287 during the period between late 2015 and 2017 shows several interesting details when viewed through the multiwavelength lens. \cite{Myserlis2018A&A...619A..88M} attributed the moderate but consistent rotation behaviour seen in the radio bands with the motion of a component travelling down the jet bend and the erratic optical subrotations with the helical movement of the jet. \cite{Lico2022A&A...658L..10L} in turn connected the passage of a moving component through a quasi-stationary feature seen in the 86 GHz VLBI images with the first VHE detection of OJ 287 in February 2017. In the following discussion, we demonstrate that by using optical data with improved cadence and including millimetre bands along with the centimetre bands we can flesh out the scenarios sketched in \cite{Myserlis2018A&A...619A..88M} and in \cite{Lico2022A&A...658L..10L}. 

One key aspect of our study was the great improvement in the cadence of the optical polarisation data that now shows a clear $\sim 210\degree$ rotation at MJD 57570 and several smaller rotations following it. This can still be explained by a knot following a spiral path in the acceleration and collimation zone (as discussed in \citealp{Myserlis2018A&A...619A..88M}, and also see \citealp{Marscher_2010arXiv1002.0806M} for a long rotation seen in $\mathrm{PKS} 1510-089$). Comparing the behaviour of the radio and the optical EVPA, it seems that the fast rotation only occurs in the optical band. In the radio bands, only at 10.45 GHz a faster portion of a rotation \citep[with a rate of $-4\degree/\mathrm{d}$ as derived in][]{Myserlis2018A&A...619A..88M} is seen starting around MJD 57450, but such a smooth rotation is not clearly seen in the other bands. As we discussed in Sect. \ref{analysis:evpa}, the mm-band data contains rather long gaps that in theory could accommodate for similar faster rotations. However, with all the radio bands (with the exception of the 2.64 GHz band) showing a similar, slow EVPA decline rate, it seems unlikely that such rotations exist in these wavelengths. \cite{Myserlis2018A&A...619A..88M} found a rotation rate of $-0.7\degree/\mathrm{d}$ for the 8.35 and 10.45 GHz bands after MJD 57500, and we confirmed similar rates for the mm-bands ($-0.8\degree/\mathrm{d}$) and the optical band ($-0.6\degree/\mathrm{d}$, after the seasonal gap). After MJD 57780, all bands stabilise to rates close to zero. After these EVPA events, the radio and optical bands have completed a full rotation of the EVPA. Since all the bands, optical included, end up in the same EVPA value (with differences of $180\degree$) this could be naturally explained with the conclusion made in \cite{Myserlis2018A&A...619A..88M} that the EVPA in both bands returns back to the stable core level after the emitting component has travelled down the bend of the jet.

\begin{figure}
\centering
\includegraphics[scale=0.4]{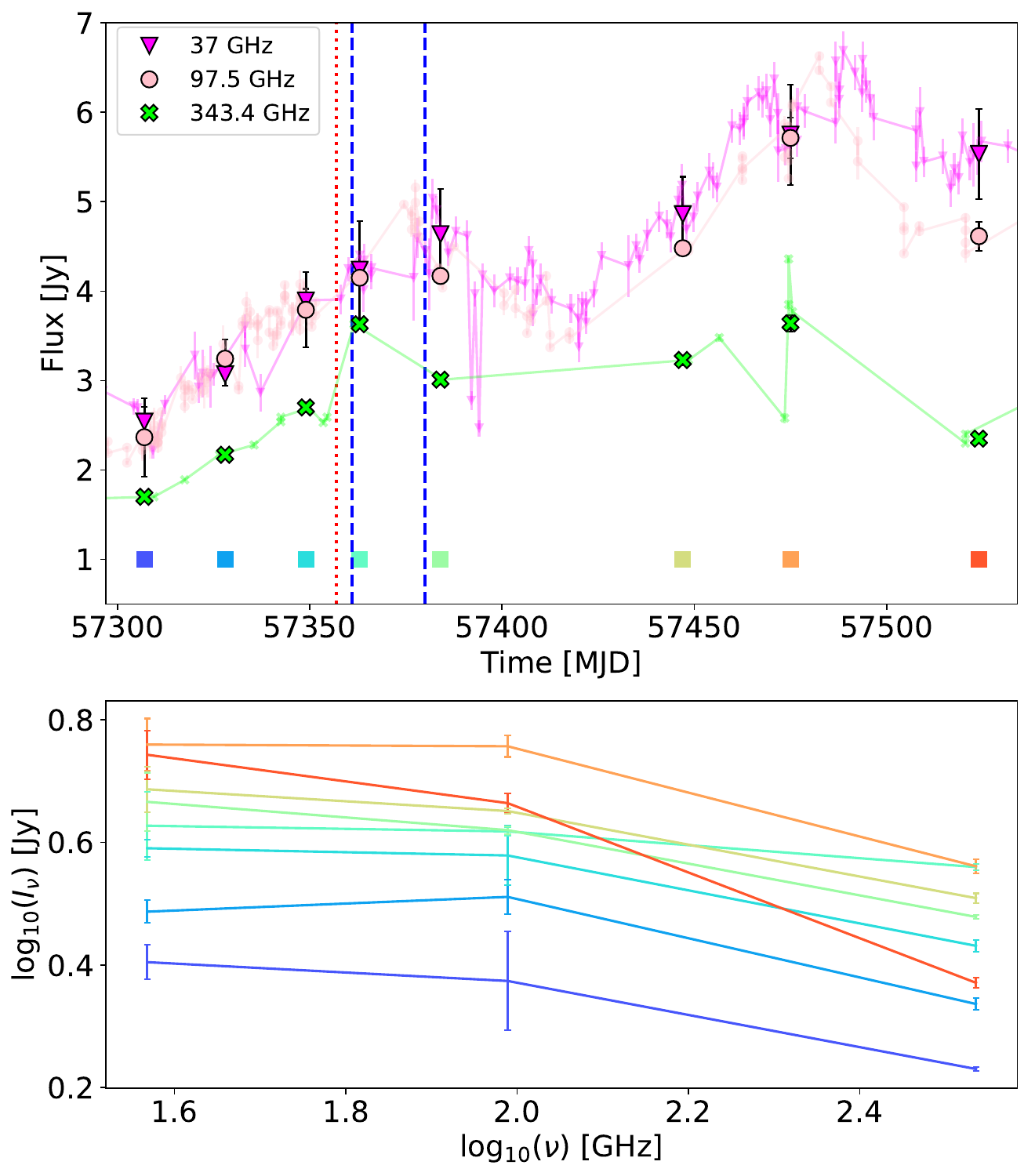}
  \caption{Top panel: light curves of 37, 97.5, and 343.4 GHz bands from the period MJD 57300-57550. To construct the spectra, the data was rebinned with a weekly cadence and averaged. Highlighted data points show the selected averaged bins of each light curve. The red dotted line shows the time when K was within the 86 GHz core. The two blue dashed lines show the times of the 2015 GR flares. Bottom panel: time evolution of the spectrum, the order of time colour-coded based on the coloured patches in the top panel.}
  \label{fig:radio_spectra}
\end{figure}

Remarkably, the start of the long optical rotation coincides with the 2015 GR flare very closely with a difference of about a week (the first GR flare peak seen on MJD 57361 by \cite{Valtonen2016ApJ...819L..37V} and the start of the rotation presented here around MJD 57369). Using the 86 GHz VLBI data presented in \cite{Lico2022A&A...658L..10L}, we calculated a new, slower speed of 0.12 mas/yr for the moving component K, and by using this revised speed we obtained the time when K should have still been merged within the 86 GHz core (around MJD 57357, see Sect. \ref{analysis:speed}). This time also coincides with the time of the 2015 GR flare and the long EVPA rotation that we see. This concurrence could suggest a link between these events that we discuss later in this section. By looking at the radio spectra in Fig. \ref{fig:radio_spectra} between the 37, 97.5, and 343.4 GHz bands, we can see a flattening in the two higher frequencies around MJD 57350. This could indicate the emergence of the moving component K from the core around that time, matching our calculation in Sect. \ref{analysis:speed}.


The passage of the moving component K through a quasi-stationary feature S1 was linked with the first VHE detection of OJ 287 in February 2017 by \cite{Lico2022A&A...658L..10L}. In our analysis, we confirmed that with the slower speed the component K was likely still interacting with the quasi-stationary feature S1 at the time of the flare. The SED modelling from the same flaring period was found to be consistent with a similar scenario where the changes in the Doppler factor of the jet blob are caused by its passage through a shocked region \citep{Acharyya2024arXiv240711848A}. Furthermore, a similar scenario was suggested for S5 0716+714 where the passage of a knot through a quasi-stationary feature was coinciding with a flaring event in the VHE gamma rays \cite{MAGIC_2018A&A...619A..45M}. This makes the scenario where such a passage could generate turbulence within the jet and be the cause of the elevated VHE activity appear very plausible.

\begin{figure}
\centering
\includegraphics[scale=0.38]{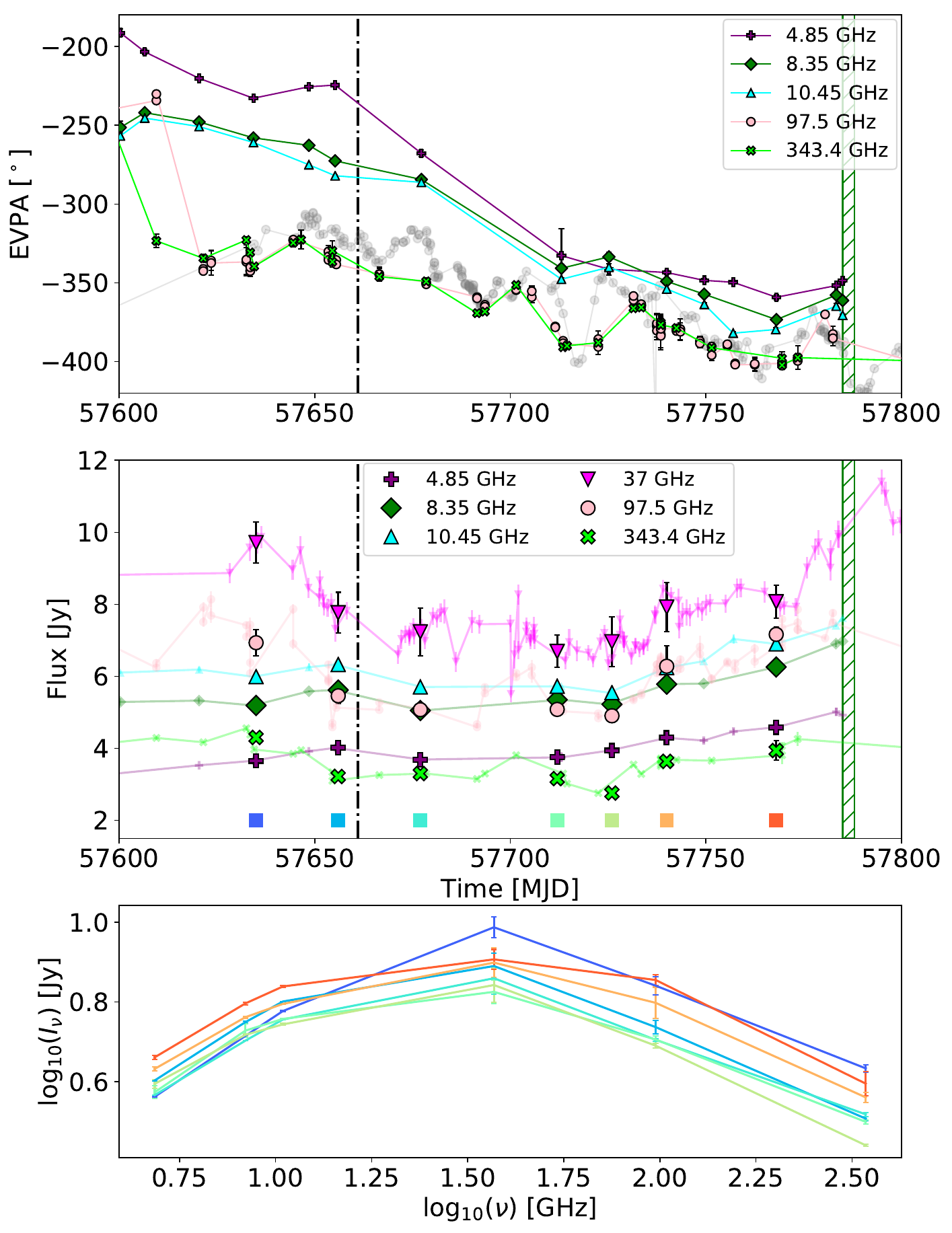}
  \caption{Top panel: EVPA curves of 4.85, 8.35, 10.45, 97.5, and 343.4 GHz. The optical data is shown in grey. The black dot-dashed line shows the epoch when S1 was detected brightened in the GMVA 86 GHz images. The green hatched area shows the epoch of the February 2017 VHE flare. Middle panel:light curves of 4.85, 8.35, 10.45, 37, 97.5, and 343.4 GHz bands from the period MJD 57600-57800. Similar to Fig. \ref{fig:radio_spectra}, the data was rebinned with a weekly cadence and averaged. Highlighted data points show the selected averaged bins of each light curve. Bottom panel: time evolution of the spectrum, the order of time colour-coded based on the coloured patches in the top panel.}
  \label{fig:evpa_zoom_spectra}
\end{figure}

Surrounding the epoch when the quasi-stationary feature S1 was seen brightening, specifically between MJD 57570-57730, there is a pronounced difference between the EVPAs of the cm-bands (4.85-10.45 GHz) and the mm-bands (97.5 and 343.4 GHz) of about a $100 \degree$ (see the top panel of Fig. \ref{fig:evpa_zoom_spectra}). The smaller differences between the cm-bands can be explained with a varying Faraday screen with rotation measures between 80-330 $\mathrm{rad/m^2}$. However, the $\sim 100 \degree$ difference between the cm- and mm-bands is much larger than that caused by similar Faraday rotation. Overall, it looks like the EVPAs in the mm-band from that epoch follow more closely the behaviour seen in the optical band, the optical being better sampled and more erratically varying. Because the mm-bands are essentially already optically thin in these jet regions that the 86 GHz also probes (see Fig. \ref{fig:evpa_zoom_spectra} bottom panel), it is possible to see the passage of K through S1 in these frequencies. In the cm-bands, these regions are still blended in together in a single core region. Therefore, a natural explanation for the differences we see in the EVPA around the time when K is residing within S1 would be the fact that the optically thin mm-bands trace these jet substructures not resolved in the cm-bands. This is in line with past observations of OJ 287 by \cite{Kikuchi_1988A&A...190L...8K}, where they observed similar, simultaneous behaviour of the optical and the radio EVPA. They suggested a connection between the optical and radio emissions although these emission likely do not come from exactly the same region, the radio emission originating from a more turbulent and less dense part of the jet.

Thus, a bigger picture takes shape. According to the GR model, the impact of the secondary BH to the accretion disk of the primary BH was seen in December 2015. We link this event with the coincident emergence of the new moving component K from the core region. Because the optical wavelengths trace the finer structure of the core still optically thick to the mm-bands, it is likely that the long  $\sim 210\degree$ optical rotation and the subsequent chaotic rotations could show the moving component K travelling down a helical magnetic field as suggested by \cite{Myserlis2018A&A...619A..88M}. All of these events coincide within two weeks, which seems incredibly fast for a new component triggered by the impact to have been able to propagate down the jet to the distances that we see. As seen by \cite{Valtonen2016ApJ...819L..37V}, the first, unpolarised flare peak is followed by a highly polarised second peak, suggesting its synchrotron origin. Such flares have been seen in the past close in time with the thermal GR flares but with much longer separations. It is then likely that any matter launched down the jet because of the impact needs at least $\sim100$ days to propagate far enough \citep{Valtaoja_2000ApJ...531..744V}. Rather, it could be that the approaching impact generates increased activity near the accretion disk and the jet base during this time, resulting in the onset of the new moving component K seen as the EVPA rotation and the spectral flattening.

A slower, general rotation trend continues in the optical band after the fast $\sim 210\degree$ rotation and is later seen in the mm- and the cm-bands in radio, as the component traverses down the bend of the jet in OJ 287. As the component K passes through the quasi-stationary feature S1, the optical- and the mm-wavelengths are both optically thin for this region, showing more variability in their EVPA curve. However, this region is not yet resolved in the cm-wavelengths, and the polarisation from different structures is blended in together, causing a difference of about $\sim 100 \degree$ in their EVPA with respect to the mm- and optical-wavelengths. In the wake of the passage of K through the shocked region S1, turbulence is created, being the likely trigger of the VHE detection from OJ 287 in February 2017 as was already suggested by \cite{Lico2022A&A...658L..10L}. After this, the moving component K has travelled down the jet bend and the EVPAs in all bands stabilise, all ending up in the same values after completing a full $360\degree$ rotation. 

\begin{acknowledgements}
J.J. was supported by the Academy of Finland projects 320085, 322535, and 345899, as well as by the Alfred Kordelin Foundation. T.H. was supported by Academy of Finland projects 317383, 320085, 322535, and 345899. E.L. was supported by the Academy of Finland projects 317636, 320045 and 346071. We thank Dr. Seiji Kameno for providing the AMAPOLA data and Dr. Alok Gupta for providing the data published in \cite{Gupta2023ApJ...957L..11G}. This publication makes use of data obtained at Metsähovi Radio Observatory, operated by Aalto University in Finland. The Liverpool Telescope is operated on the island of La Palma by Liverpool John Moores University in the Spanish Observatorio del Roque de los Muchachos of the Instituto de Astrofisica de Canarias with financial support from the UK Science and Technology Facilities Council. The research at Boston University was supported in part by the National Science Foundation grant AST-2108622, and a number of NASA Fermi Guest Investigator grants,
the latest is 80NSSC23K1507.
This study was based in part on observations conducted using the 1.8m Perkins Telescope Observatory (PTO) in Arizona, which is owned and operated by Boston University. Data from the Steward Observatory spectropolarimetric monitoring project were used. This program is supported by Fermi Guest Investigator grants NNX08AW56G, NNX09AU10G, NNX12AO93G, and NNX15AU81G.
\end{acknowledgements}

%
%

\bibliographystyle{aa} 
\bibliography{OJ287_paper} 


\end{document}